\documentclass[conference]{IEEEtran}
\IEEEoverridecommandlockouts
\usepackage{cite}
\usepackage{amsmath,amssymb,amsfonts}
\usepackage{graphicx}
\usepackage{textcomp}
\usepackage{xcolor}
\usepackage{algorithm}
\usepackage{algpseudocode}

\def\BibTeX{{\rm B\kern-.05em{\sc i\kern-.025em b}\kern-.08em
    T\kern-.1667em\lower.7ex\hbox{E}\kern-.125emX}}
\begin{document}

\title{Hybrid Quantum Tabu Search for \\ Solving the Vehicle Routing Problem}

\author{\IEEEauthorblockN{James B. Holliday}
\IEEEauthorblockA{\textit{EECS Department} \\
\textit{University of Arkansas}\\
\textit{J.B. Hunt Inc.}\\
Fayetteville, AR, USA \\
jbhollid@uark.edu}
\and
\IEEEauthorblockN{Braeden Morgan}
\IEEEauthorblockA{\textit{EECS Department} \\
\textit{University of Arkansas}\\
Fayetteville, AR, USA \\
braedenm@uark.edu}
\and
\IEEEauthorblockN{Hugh Churchill}
\IEEEauthorblockA{\textit{Dept. of Physics} \\
\textit{University of Arkansas}\\
Fayetteville, AR, USA \\
hchurch@uark.edu}
\and
\IEEEauthorblockN{Khoa Luu}
\IEEEauthorblockA{\textit{EECS Department} \\
\textit{University of Arkansas}\\
Fayetteville, AR, USA \\
khoaluu@uark.edu}
}

\maketitle

\begin{abstract}
There has never been a more exciting time for the future of quantum computing than now. Real-world quantum computing usage is now the next XPRIZE. With that challenge in mind we have explored a new approach as a hybrid quantum-classical algorithm for solving NP-Hard optimization problems. We have focused on the classic problem of the Capacitated Vehicle Routing Problem (CVRP) because of its real-world industry applications. Heuristics are often employed to solve this problem because it is difficult. In addition, meta-heuristic algorithms have proven to be capable of finding reasonable solutions to optimization problems like the CVRP. Recent research has shown that quantum-only and hybrid quantum/classical approaches to solving the CVRP are possible. Where quantum approaches are usually limited to minimal optimization problems, hybrid approaches have been able to solve more significant problems. Still, the hybrid approaches often need help finding solutions as good as their classical counterparts. In our proposed approach, we created a hybrid quantum/classical metaheuristic algorithm capable of finding the best-known solution to a classic CVRP problem. Our experimental results show that our proposed algorithm often outperforms other hybrid approaches.
\end{abstract}


\section{Introduction}

The Vehicle Routing Problem (VRP) is an NP-hard problem first introduced in 1954 \cite{dantzig1954solution}. Since then, it has been widely explored in numerous research papers \cite{eksioglu2009vehicle}. The VRP generalizes another NP-hard problem famously known as the Traveling Salesman Problem (TSP). Where the TSP attempts to find the shortest path to visit \(N\) locations on a single tour or route, the VRP breaks the problem up using one or more routes to visit all the locations. Reference \cite{eksioglu2009vehicle} points out that it was actually in a 1964 paper \cite{clarke1964scheduling} that multiple vehicles were first considered to solve the problem, which is how we understand the VRP today. Both problems are still widespread in the transportation industry, and companies solve them numerous times daily as part of their operations.

\begin{figure}[t]
    \centering
    \includegraphics[width=\linewidth]{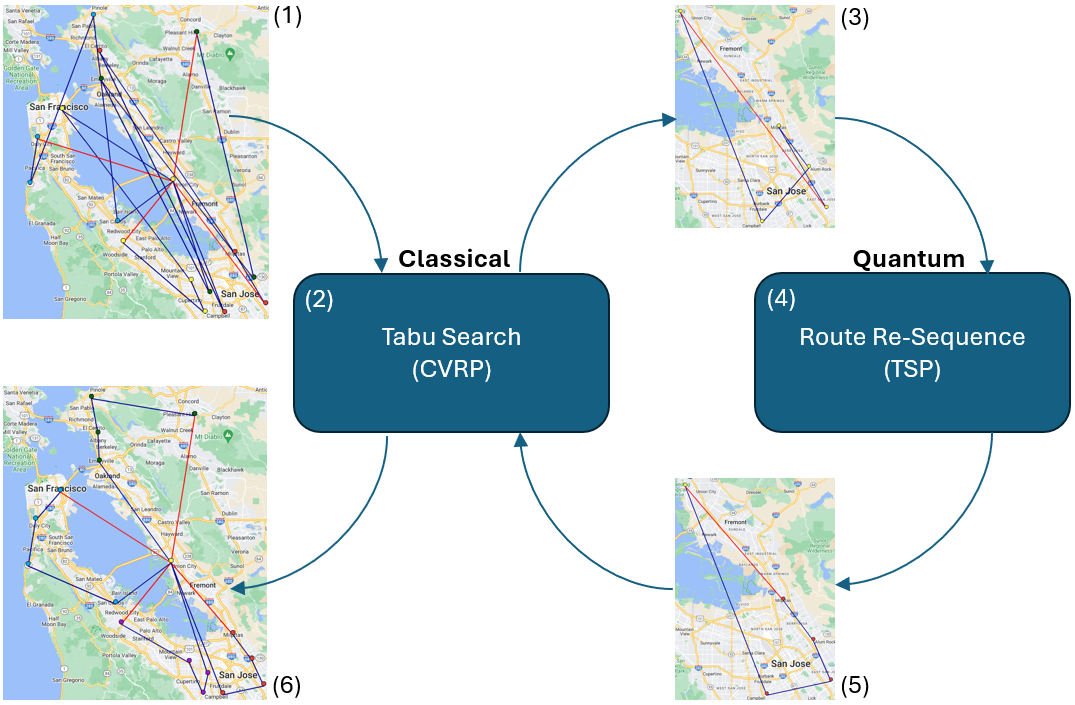}
    \caption{Our Proposed Approach. Step 1. An initial solution is provided to the classical TS algorithm. Step 2. TS is performed. Step 3. At different times during TS an individual route from a solution is selected to be optimized. Step 4: Route re-sequencing (TSP) is performed via QUBO formulation on a QA. Step 5. The re-sequenced route is returned to TS. Step 6. When TS reaches its stopping criteria, the best solution found is returned.}
    \label{fig:proposed}
\end{figure}

Quantum computing (QC) \cite{ladd2010quantum} is a critical emerging technology promising to solve NP-hard problems like the VRP much faster than classical computers. Whereas with classical computing, solving for optimal solutions is often intractable due to the time it takes. Heuristics and metaheuristics \cite{laporte2013vehicle} are usually used to solve the VRP because they are faster and provide near-optimal results. Quantum computers do not currently contain enough qubits to solve large and complicated variants of the VRP \cite{irie2019quantum} in a single encoded problem. Limited numbers of qubits are the primary challenge facing recent quantum computers, so how can we utilize the currently available quantum computers to aid in solving NP-hard problems like the VRP? Hybrid algorithms are able to bridge this gap and allow us to utilize the quantum resources that are available now, perhaps to improve our classical approaches \cite{feld2019hybrid}.

Hybrid quantum-classical algorithms are an emerging trend in research. In a 2022 survey on this topic \cite{osaba2022systematic}, it shows that half of the papers focused on solving the VRP or TSP utilized hybrid approaches. Of those hybrid approaches, seventy-five percent (12 of 16) utilized the Quantum Annealer (QA) type of quantum computer. A QA solves optimization problems by taking advantage of the properties of quantum mechanics. Using quantum fluctuations to guide the optimization process towards the optimal solution. These fluctuations are the energy states of a problem, and the annealer attempts to find the lowest-energy state that is also the optimal combination of the problem elements \cite{dwave2024}. Hybrid algorithms overcome quantum computing's limitations on problem size and variable count by breaking the optimization problem into smaller parts and optimizing some parts using classical computing and other parts using quantum computing. In order to utilize the QA, the problem must be set up in a way that the QA can process. The formulation we have chosen is the Quadratic Unconstrained Binary Optimization (QUBO) problem. Using a hybrid approach the entire optimization problem needs not be formulated into the QUBO, thus keeping the QUBO small enough to be solved by a QA.  

\textbf{Our Contributions in this Work:} 
In this paper, we focus on the research of hybrid algorithms to solve combinatorial optimization problems. Due to its practical applications in industry, this research focuses on a widespread variant of the VRP, the Capacitated Vehicle Routing Problem (CVRP). We propose the first hybrid quantum-classical metaheuristic algorithm to solve the VRP. The algorithm is a metaheuristic that interacts with a QA at different times throughout the search. The classical portion of the algorithm is used to make sure constraints are not broken and to perform the local search. The quantum portion is used to optimize the routes discovered during the search.  

The rest of this paper is organized in the following format. Section \ref{Sec:Background} will review the background. Section \ref{Sec:ProposedMethod} will introduce our proposed method. Section \ref{Sec:Results} will present the experiments, and Section \ref{Sec:Concl} will be our conclusions.

\section{Background} \label{Sec:Background}

\subsection{Vehicle Routing Problem}
The Vehicle Routing Problem is the problem of finding the optimal, i.e., shortest, route or collection of routes that visit a set of delivery locations, cities, or customers such that no constraints are broken, meaning that each route is feasible. A route is the sequence where each location is visited. For our version of the VRP, a route must start and stop at a depot from which a vehicle will travel to each location along the route.  

The VRP formally defined is an NP-hard combinatorial optimization problem with a directed graph \(G = (V, E)\), where \(V = \{v_0,...,v_n\}\) is the set of locations and \(E = \{ (v_i,v_j): v_i,v_j\in V, i \ne j)\}\) is the set of edges between the locations. Location \(v_0\) is not a delivery location but is the depot where the routes begin and end. The set or fleet of vehicles is \(K\). We will discuss constraints around vehicles in the next section. There is also a cost matrix over the edges \(C = (c_{ij})\). We define cost as the distance between two locations in the graph. In order to solve the VRP, we must create at most \(|K|\) routes that start and end at the depot location and minimize total cost. For simplicity, we will say the set \(N\) is equal to \(V\), except it does not contain the depot, \(v_0\). \(
x_{ijk}\) is defined as the binary variable \(x\) for an edge between location \(i\) and location \(j\) on vehicle \(k\). We can mathematically represent this problem as a minimization problem with constraints as follows:
\begin{align}
 & min \sum_{k\in K} \sum_{i\in V} \sum_{j\in V} c_{ijk} x_{ijk}
 \end{align}
 s. t. 
 \begin{align}
 & \sum_{k\in K} \sum_{j\in V} x_{ijk} = 1 \quad \forall i \in V\\
 & \sum_{j\in N} x_{0jk} = 1 \quad \forall k \in K\\
 & \sum_{i\in N} x_{i0k} = 1 \quad \forall k \in K\\
 & \sum_{i\in N} x_{ijk} - \sum_{i\in N} x_{jik} = 0 \quad \forall j \in N, k \in K\\
 & 2 \leq u_{ik} \le N \quad \forall i \in N, k \in K\\ 
 & u_{ik} - u_{jk} + 1 \leq (N - 1)(1 - x_{ijk}) \quad \forall i, j \in N, k \in K\\
 & x_{ijk} \in \{0,1\}
\end{align}
In this equation, (1) is the minimization function subject to additional constraints. Constraint (2) ensures that each location is visited by only one vehicle. Constraints (3) and (4) ensure that each route on each vehicle starts and ends at the depot. Constraint (5) ensures that the number of entries into a location equals the number of exits. Constraints (6) and (7) are necessary to prevent sub-tours. Constraint (8) defines the primary binary decision variable.  


\subsection{Variants of the Vehicle Routing Problem}

There are numerous variations of the VRP \cite{eksioglu2009vehicle}. VRP with Time Windows (VRPTW) and the CVRP are the most similar to the variants solved by the transportation industry and have the most academic datasets available. We have focused on one of the most well-studied variants, the CVRP. The primary differentiation from the VRP for the more constrained CVRP is that now each vehicle has a limited capacity \(Q\), so the sum of each location's demand \(q\) along the route must be less than or equal to the vehicle's capacity. In this preliminary study, we primarily focus on the homogeneous fleet variant of the problem where each vehicle has the same value for \(Q\). A heterogeneous fleet variant (HVRP) also allows each vehicle to have a different value for \(Q\). Here is the additional mathematical formulation for the CVRP as shown in Eqn. \eqref{eqn:xjik}.

\begin{equation} \label{eqn:xjik}
 \sum_{i\in V} q_i \sum_{j\in V} x_{ijk} \leq Q \quad \forall k \in K\\
\end{equation}

The constraint in Eqn. \eqref{eqn:xjik} limits the customers on a route so that the capacity of the vehicle is not exceeded. This constraint is added to the list of constraints from the previous section to define the CVRP fully.    

\subsection{Adiabatic Quantum Computing}
While classical computers rely on bits, quantum computers leverage quantum bits, or qubits, to exploit the principles of quantum mechanics. Unlike classical bits restricted to either 0 or 1, qubits can exist in a superposition of both states simultaneously. When multiple qubits are combined, the collective quantum state, denoted by $|\psi(t)\rangle$, is governed by the Schrödinger equation as shown in Eqn. \eqref{eqn:par}.

\begin{equation} \label{eqn:par}
i \frac{\partial}{\partial t} |\psi(t) \rangle = H(t) |\psi(t) \rangle
\end{equation}

In Eqn. \eqref{eqn:par}, the Hamiltonian, \(H(t)\), is an operator that dictates the system's energy at a specific time, \(t\). For problems encoded as Quadratic Unconstrained Binary Optimization (QUBO) problems, the solution lies in the state of the qubits when the QUBO Hamiltonian, $H_{\text{QUBO}}(t)$, reaches its lowest energy state. Constant \(i\) is the imaginary unit. However, directly initializing the system in this minimal energy state for a QUBO problem proves challenging. To circumvent this hurdle, researchers often employ the strategy of Adiabatic Quantum Computing. This approach leverages a readily computable Hamiltonian, $H_{\text{BASIC}}(t)$. With the system in the ground state of $H_{\text{BASIC}}(t)$), we utilize an interpolation between the two Hamiltonians to construct the time-dependent Hamiltonian:

\begin{equation}
H(t) = \left( 1 - \frac{t}{T} \right) H_{\text{BASIC}}(t) + \frac{t}{T} H_{\text{QUBO}}(t)
\end{equation}

The Adiabatic Theorem guarantees that if the variation of $0 \leq t \leq T$ is sufficiently slow, the system's state will remain in the ground state throughout the interval \([0, T]\). Consequently, measuring the quantum state at time \(T\) will yield a solution that satisfies the QUBO problem, as it will be in its minimal energy state for $H_{\text{QUBO}}(t)$ \cite{farhi2000quantum}.

There have been numerous research efforts in solving the entire VRP and some of its variants on quantum devices. Some recent research studies in this field are \cite{harwood2021formulating}, which created multiple QUBO formulations for the VRPTW. Their research did not attempt to solve large-scale versions of the problem but instead focused on the complexity of their solution. Research by \cite{fitzek2021applying} focused on the HVRP that could be solved with 11, 19, and 21 qubits. They simulated QC using classical optimizers. An application example of the VRP was mentioned in \cite{matsubara2020digital}, and they stated QC could solve a VRP with approximately 30 depots and a vehicle count of 48. The number of locations visited on the routes was not provided making it hard to compare to other solutions. Still it showcases one of the larger problem sizes in recent research. However, the formulation for their solution was not provided. While these efforts show the promise of large-scale QC's ability to solve the VRP, except for one, they do not show a near-term quantum ability to solve significant real-world versions of the problem, which is the focus of our research.  
\subsection{Hybrid Quantum Algorithms}
Our research review discovered other hybrid approaches to solving the VRP. Reference \cite{harikrishnakumar2020quantum} focused on formulating a single QUBO to solve the Multi-Depot Capacitated Vehicle Routing Problem (MDCVRP), and they did not present any results. Their approach was only considered hybrid because of DWAVE's ability to split large QUBOs into smaller QUBOs so that it can map onto the system's available qubits. Similarly, \cite{yarkoni2021solving} solved the Shipment Rerouting Problem, with problems up to 100 locations, in a hybrid fashion like \cite{harikrishnakumar2020quantum}. Reference \cite{bao2021approach} focused on solving the Vehicle Routing Problem with Balanced Pick-up (VRPBP). Reference \cite{bao2023ising} are the same authors and expanded the work in \cite{bao2021approach}. They created both 2-phase and 3-phase approaches where all phases are formulated as QUBO's, and they presented results in their research. Lastly, an interesting cost function was defined in \cite{ajagekar2020quantum}, where a ratio between distance and work time was used to define the cost. They showed the results of their hybrid algorithm for problems of up to twelve locations.    

Reference \cite{feld2019hybrid} is the most highly cited hybrid approach. It is a hybrid algorithm that operates as a 2-phase heuristic with clustering and routing phases. In order to create the clusters, they use a clustering core point, which can be set as a location with the most significant demand or the location farthest from the depot. From there, they perform a clustering algorithm to add additional locations to the cluster until it reaches vehicle capacity. This process is repeated until all locations are clustered. They then perform a cluster improvement method where locations are moved between clusters if the distance to the cluster center is reduced by making a move. This process is repeated for several iterations, or no moves are possible. 

Once phase one is complete, the second phase of the routing phase of the heuristic is performed, where the clusters are formulated as a QUBO for solving the TSP as defined in \cite{lucas2014ising}. Here we alter the notation from \cite{lucas2014ising} and continue with (\(ij\)) defined as before as the edge set from location \(i\) to location \(j\), and \(u, v\) are the sequence on the route that a location is visited. Meaning \(x_{i,u}\) is the binary variable that represents location \(i\) is visited as the \(u\)th stop on the route.  \(n\) contains all the locations being routed and \(N\) is equal to \(|n|\).

\begin{multline}\label{eq12}
 H_A = A \sum_{j = 1}^n \left( 1 - \sum_{v = 1}^N x_{j,v}\right)^2 + A \sum_{v = 1}^n \left(1 - \sum_{j = 1}^N x_{j, v}\right)^2 + \\ A \sum_{ij \notin E} \sum_{v = 1}^N x_{i,v}x_{j,v+1}
\end{multline}

Here Eqn. \eqref{eq12} is the QUBO formulation for the Hamiltonian Cycle Problem. The first term ensures that every location appears in the cycle. The second term ensures a $v$th node in the cycle for each $v$. The third term ensures an edge must exist from $i$ to $j$.

\begin{equation} \label{eqn:HB}
 H_B = B \sum_{ij \in E} C_{ij} \sum_{v = 1}^{N}x_{i,v}x_{j,v+1}
\end{equation}

Eqn. \eqref{eqn:HB} ensures the cost of the Hamiltonian Cycle is minimized. \(C_{ij}\) again is the cost to travel from location \(i\) to location \(j\).

\begin{equation}
 H = H_A + H_B
\end{equation}

So adding both Hamiltonians together in (14) is the QUBO that will solve the TSP or what we refer to as the route re-sequencing problem and what \cite{feld2019hybrid} referred to as the routing phase problem. The penalty coefficients are set with \(A\) being higher than the most significant cost in \(C\) and \(B\) being set to 1.  

Reference \cite{borowski2020new} is another hybrid algorithm that operates as a 2-phase heuristic. Here, the clustering is done using recursive-DBSCAN clustering, and again, sequencing is done in the same way as \cite{feld2019hybrid}. Both \cite{feld2019hybrid} and \cite{borowski2020new} claim success from their experiments, but \cite{borowski2020new} failed to explain their results in a way that is comparable to other research. Both papers ran experiments using the dataset from \cite{christofides1981exact}. Reference \cite{borowski2020new} did provide source code, so we recreated their results.   

\subsection{Metaheuristic Optimization}
Using a heuristic algorithm can often lead to a "good enough" solution to the CVRP, and heuristics can usually find that solution quickly. What is gained in solution completion time could be improved in solution quality. In the transportation industry, companies have often found that getting close quickly can still be helpful when making a recommendation. These recommendations would be provided to a logistic planner, who may manually adjust the solution before generating the final load plan for the routes. Generally, the transportation industry is willing to make this trade-off as long as they know their recommendation systems are creating a near-optimal solution. Many heuristics have been providing solutions that make this trade-off for a long time. Clark and Wright's 1964 savings algorithm \cite{clarke1964scheduling}, while perhaps the oldest heuristic for the CVRP, is still one of the fastest and is capable of obtaining near-optimal solutions for some problems. References \cite{gillett1974heuristic} and \cite{fisher1981generalized} are also well-known heuristics for the CVRP, and we report our results compared to all three of these algorithms.  

Is there an alternative to heuristics that can provide better solutions? What other options exist to solve the CVRP? Metaheuristics are algorithms that combine search algorithms to find better solutions across the space of all solutions. Usually they use a combination of different heuristics to create neighborhoods within the search space to perform local search then on. There are many different types of metaheuristics and many different ways to categorize them. Famous examples include the genetic algorithm, ant colony optimization, and simulated annealing. As stated in \cite{cordeau2002guide}, a type of metaheuristic named Tabu Search (TS) has been shown to outperform other metaheuristics, especially concerning the CVRP. In many classical CVRP datasets, TS has found the best-known solution (BKS).    

While \cite{osaba2021hybrid} did not solve the VRP with their hybrid algorithm, they implemented TS to solve the TSP.

\subsection{Tabu Search}
TS was formally introduced in \cite{glover1989tabu} and \cite{glover1990tabu}. TS is a type of local search algorithm. A local search moves from a current solution to a better solution by choosing a move in the current solution's neighborhood. A neighborhood is defined by any solution adjacent to the current solution by making a simple change to the current solution. In the case of the CVRP, a local search move could be defined as moving a stop location from one route to another. In TS, the neighborhoods are created by attempting every possible move on the current solution and choosing the move that leads to the best new solution. There are TS implementations that do not always consider every possible move but stop looking for the best move as soon as any improving move is found \cite{osman1993metastrategy}. The process is iterative in that the new best solution becomes the current solution, and a new neighborhood is generated again. Using a short-term memory component called the tabu list allows this local search concept to avoid becoming stuck in a local optimum. A move is marked tabu and added to the tabu list when that move was the move that created the new best solution. Move's are kept on the tabu list for several iterations before they can be used again. This tabu list sometimes forces the best solution to be worse than the current solution and forces the local search into new parts of the global solution space.

According to \cite{osman1993metastrategy}, any TS implementation must define four strategies. First, "forbidding," which is used to mark some moves as tabu. Second, the "freeing" strategy is used to un-mark moves previously tabu as no longer tabu. Third, the "short-term" strategy controls the first two strategies. Additionally, this strategy should define two more strategies; the "aspiration" strategy on when it is permissible to ignore that a move is marked as tabu, and the "selection" strategy, which determines if every possible move should be checked or just the first improving move. Lastly, the "stopping" strategy controls when the TS should end its search. In our method, we utilize many of these concepts defined in \cite{osman1993metastrategy} and other "strategies" we will present in the next section.

The first step in TS is to construct a starting solution to start a local search. There are many different ideas on how to construct the starting solution. Reference \cite{osman1993metastrategy} proposed using the savings algorithm from \cite{clarke1964scheduling}. While \cite{taillard1993parallel} discussed using random assignment to generate the starting solution.

Once the starting solution is created, iterations can begin where a local search is performed to find better and better solutions. The "forbidding" and "freeing" strategies are implemented here. Moves are made tabu and later made not tabu as the process runs in a loop. Many different types of moves have been considered, and \cite{osman1993metastrategy} does an excellent job explaining some of them. A (1,0) or (0,1) move is defined as moving a location from one route to another. A (1,1) is a swap where locations are exchanged between two routes. Where \cite{osman1993metastrategy} utilized (1,0), (0,1), and (1,1) moves, \cite{taillard1993parallel} claims they only performed (1,0) moves. Still other's utilized more complicated swapping moves like GENI in \cite{gendreau1994tabu}'s TABUROUTE. GENI \cite{gendreau1992new} stands for generalized insertion routine. It only allows locations to be inserted into a route if it contains one of its closest neighbors. Insertions are executed simultaneously with a local re-optimization of the route's sequence.        

\begin{figure}[t]
    \centering
    \includegraphics[width=\linewidth]{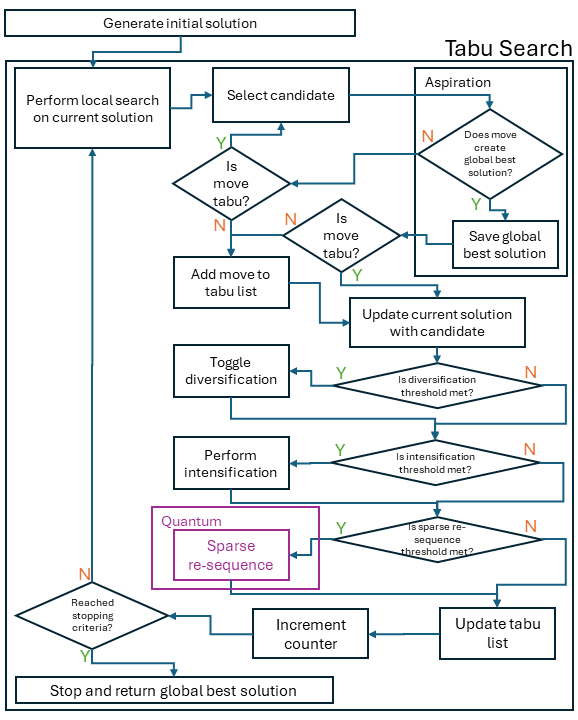}
    \caption{Flowchart of Tabu search w/o strategic oscillation}
    \label{fig:flowchart}
\end{figure}

\section{Our Proposed Method} \label{Sec:ProposedMethod}

According to \cite{cordeau2002guide}, the best TS implementations can outperform simulated and deterministic annealing, genetic search, ant systems, and neural networks. With that motivation, we created a hybrid quantum/classical TS implementation. The goal is to show that by utilizing hybrid algorithms, quantum computing in the near term can provide value even while qubit quantity is limited. We compartmentalize our algorithm into parts following from \cite{feld2019hybrid} and \cite{borowski2020new}. The first part is the tabu search algorithm, which we perform as a classical algorithm, and the second is the route sequencing algorithm, which we convert to QUBO form and solve using D-WAVE's quantum annealer. Together, we call this Hybrid Quantum Tabu Search (HQTS).

\subsection{Hybrid Quantum Tabu Search}

We implement many standard TS enhancements for our method, such as aspiration, intensification, and diversification \cite{glover1989tabu}. TS comprises the following parts: create a starting solution, local search for a new solution, update moves in the tabu list, perform search enhancements, and sparsely perform route re-sequencing; if the stopping criteria have not been met, a new local search is performed. 

\subsection{Aspiration}

After creating the starting solution, the rest of the algorithm runs inside a loop. The first step in the loop is to search locally for a move that will create the next solution. If the move leads to a global best solution, that move is kept regardless of whether it is marked tabu or not, which is how aspiration is defined by \cite{osman1993metastrategy}. If a new global best solution is not found, the best non-tabu move is selected; note that this might lead to a lower quality solution than the current one. The process is repeated until a stopping criterion is reached.  

\subsection{Local Search}

In order to set up our starting solution, we first grouped the locations similarly to \cite{feld2019hybrid}. For each problem we experimented with, we knew a BKS and the number of vehicles used in that BKS. If a BKS was unknown for an experiment, we used the number of vehicles found when using [16] hybrid solution. The grouping was set up so that each \(v\) in \(V\) determines its \(K\) nearest neighbors where \(K\) is the number of vehicles in the BKS. From there, the locations with the most significant values in the cost matrix \(c\) were assigned to a route so long as they were not the nearest neighbor to a location already placed on a route for \(K + 1\) vehicles. We used \(K + 1\) to ensure all the locations could be routed in the starting solution. After the routes were seeded, each route was filled by iterating through the remaining locations sorted by highest demand to lowest. The only rule was that for a location to be added to a route, at least one nearest neighbor to that location must exist. If a location could not be added to a route with a neighbor because it would exceed the vehicle capacity any remaining route with space was assigned. The motivation behind this method was that the seed locations should be relativity far apart and representative of locations that would not ultimately be routed together in the final solution. The hope was that neighboring locations would be routed together in the starting solution.  

With the starting solution created, we start TS. We continue to utilize the neighbor concept here as we build the neighborhood of moves or candidate solutions (we use these terms interchangeably) that will lead to the next solution. When a move is considered, we still hold that the route a location is being added to must contain at least one neighbor of that location. It speeds the search process by not allowing a large number of moves. The neighborhood is constructed by performing either a (1,0), (0,1) or (1,1) moves. For our method, we define the (0,1) as swapping a location with another location on the same route. We also define the (1,0) move as you would assume, but the location sequence in the new route is determined by evaluating which spot in the sequence leads to the lowest cost route, as was used in \cite{taillard1993parallel}. (1,1) was performed as described in section 2. Once all possible moves have been generated, each is evaluated by calculating the total cost of the solution each move creates. As we described before, the selected move is used to create a new solution from the neighborhood. This new solution becomes the basis as we start the local search again.  

\subsection{Intensification and Diversification}
If during iterating, the global best solution is not improved within \(X\) iterations, we initiate a diversification process where the \(K\) nearest neighbors are adjusted to be \(2 * K\) nearest neighbors. The search continues for another \(X\) non-improving move, and intensification is performed. Intensification forces the following solution to search from to be the globally best-found solution found so far in the process. After intensification, other \(X\) non-improving moves are allowed before diversification is disabled and the \(K\) nearest neighbors. \(X\) is selected uniformly between 0.6 * V and 1.1 * V. These values were determined by experimentation. Additionally, when diversification is turned on, we stop the (0,1) moves from being considered in order to speed up the search since more locations will be considered by the adjusted \(2 * K\)  nearest neighbors.   

\subsection{Sparse Re-sequencing}
Reference \cite{gendreau1994tabu} proposed re-sequencing routes on every considered move. While this speeds up the search, it is computationally expensive. Because of this cost \cite{taillard1993parallel} only performed a true re-sequence of the routes every 200 iterations. We follow from \cite{taillard1993parallel} in limiting how often a re-sequence is performed because we use a QA to perform this step. For our research, access to the QA was limited. With this being the case, a sparse re-sequence was implemented. Before a route is re-sequenced, we check a dictionary of previously optimized routes and do not call the QA if the route is found in the dictionary; this idea was also used in \cite{osaba2021hybrid}. In order to perform the re-sequence, the route and optimization problem must be formulated as a QUBO. We use the same formulation as is described in section 2.4. The QUBO is then sent to DWAVE's cloud quantum computing API, and a result is returned. The result is translated back into a route and is stored in the dictionary. Then, the re-sequenced route replaces the route in the solution. For our experiments, we performed sparse re-sequencing on the global best solution if TS still needed to improve its global best solution in 1,000 iterations.    

\subsection{Strategic Oscillation}
Forcing the search to stay in the feasible solution space has been shown to limit search effectiveness \cite{glover2011case}. In addition to the proposed method we have described thus far, we also experimented with a mechanism that can allow the search to visit infeasible space called strategic oscillation (SO) \cite{glover1989tabu}. We implemented a simple version of SO in Algorithm \ref{alg:cap} from \cite{glover2011case}. To summarize Algorithm \ref{alg:cap}, if the last candidate selected was feasible at each iteration, the evaluation for the candidates is kept the same in that the best non-tabu move is selected. However, we no longer force each move to maintain feasibility. If the best improving non-tabu move breaks feasibility that move can also be selected. Once we cross over into infeasibility, the evaluation of moves changes to no longer select the most improving move but instead to choose the best move that improves the infeasibility measure or causes the infeasibility to be worsened the least. We show in section 4 the results with and without SO. Additionally, for the SO-enabled experiments, we did not use the intensification defined in 3.4 but instead relied only on the sparse re-sequencing as a means of intensification.

\begin{algorithm}[hbtp]
\caption{Candidate evaluation with strategic oscillation}\label{alg:cap}
\begin{algorithmic}
\Require {$cbs$ can only be assigned a feasible solution so that it can be evaluated as a global best solution}
\State $p \gets \textrm{previously selected solution}$
\State $N \gets \textrm{candidate solutions from local search}$
\State $n \gets \textrm{number of candidate solutions}$
\State $cbs \gets unassigned$   \Comment{current best solution}
\State $sbfs \gets unassigned$  \Comment{selected best feasible solution}
\State $sbis \gets unassigned$  \Comment{selected best infeasible solution}
\State $ss \gets unassigned$    \Comment{selected solution}
\If{$IsFeasible(p)$}
    \For{$i \gets 1 \textrm{ to } n$}
        \If{$Cost(N_i) < Cost(sbfs) \textrm{ and } IsFeasible(N_i)$}
            \If{$Cost(N_i) < Cost(cbs)$}
                \State $cbs \gets N_i$      
            \EndIf
            \If{$!IsTabu(N_i)$}
                \State $sbfs \gets N_i$
            \EndIf
        \ElsIf{$Cost(N_i) < Cost(sbis) \textrm{ and } !IsFeasible(N_i)$}
            \If{$!IsTabu(N_i])$}
                \State $sbis \gets N_i$
            \EndIf
        \EndIf
    \EndFor
    \If{$Cost(sbis) < Cost(sbfs$}
        \State $ss \gets sbis$
    \Else
        \State $ss \gets sbfs$
    \EndIf 
\Else
    \For{$i \gets 1 \textrm{ to } n$}
        \If{$Infeasibilty(N_i) < Infeasibilty(sbfs) \textrm{ and } IsFeasible(N_i)$}
            \If{$Cost(N_i) < Cost(cbs)$}
                \State $cbs \gets N_i$
            \EndIf
            \If{$!IsTabu(N_i)$}
                \State $sbfs \gets N_i$
            \EndIf
        \ElsIf{$Infeasibilty(N_i) < Infeasibilty(sbis) \textrm{ and } !IsFeasible(N_i)$}
            \If{$!IsTabu(N_i)$}
                \State $sbis \gets N_i$
            \EndIf
        \EndIf
    \EndFor
    \If{$Infeasibilty(sbis) < Infeasibilty(sbfs)$}
        \State $ss \gets sbis$
    \Else
        \State $ss \gets sbfs$
    \EndIf 
\EndIf

\end{algorithmic}
\end{algorithm}

\subsection{Stopping Criteria}

If no new best solution has been discovered in the last 5000 moves, stop, or if 60 minutes have elapsed.

\section{Experiments and Results} \label{Sec:Results}

This section outlines the experimental setup used to evaluate the performance of the proposed hybrid quantum approach to solve the vehicle routing problem.

The algorithms that were used as a benchmark against our implementation of HQTS include the Density-Based Spatial Clustering of Applications with Noise (DBSS) and the Solution Partitioning Solver (SPS),  both of which are publicly available at \cite{githubdwave}. These algorithms were introduced in \cite{borowski2020new} and are considered hybrid quantum algorithms using QA for route optimization. From here on we will refer to HQTS as just TS.

\subsection{Classic Dataset}
\begin{figure}[t]
\centering
\fbox{\includegraphics[width=0.9\linewidth]{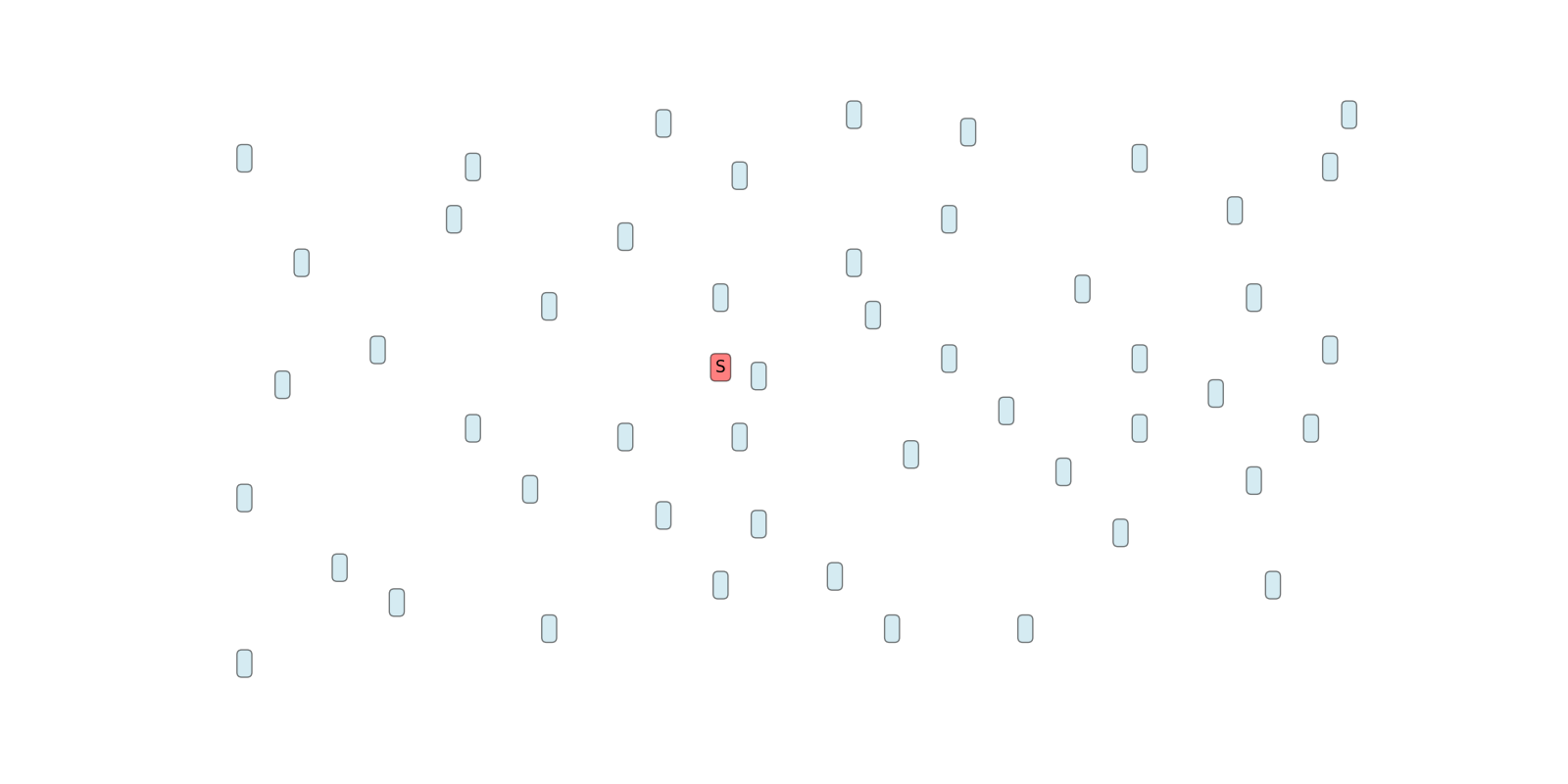}}
\caption{Visualization of CMT 1}
\label{fig:cmt1visualization}
\end{figure}

\begin{figure}[t]
\centering
\fbox{\includegraphics[width=0.9\linewidth]{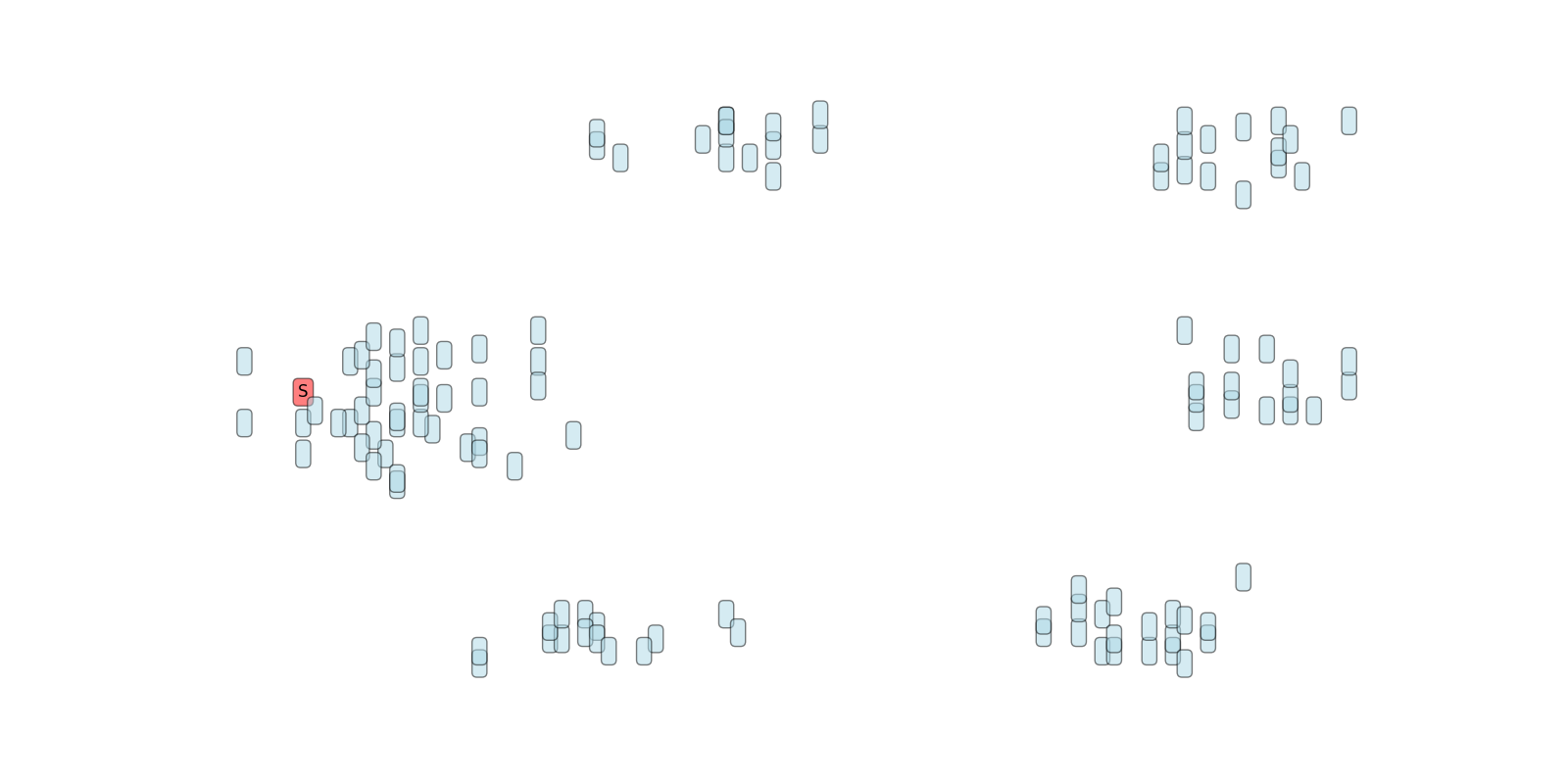}}
\caption{Visualization of CMT 11}
\label{fig:cmt11visualization}
\end{figure}

First, we evaluated the performance on seven standard CVRP benchmark datasets: \cite{christofides1981exact} (CMT) 1-5, 11, and 12. This dataset was chosen as the benchmark because it is well-established and well-studied. Additional \cite{feld2019hybrid} also reported their results against this dataset. The dataset originated in 1979 and consists of 14 problems. Based on a Belgium road network, these problems provide an academic standard for testing vehicle routing algorithms. Each problem is comprised of three sections. The first section outlines the problem. It includes the name, Best Known Solution (BKS), dimensions, vehicle capacities, and distances. The capacity is the same for each vehicle. The second section consists of nodes; each is numbered, and their coordinates are given on a 2d M64 dimensional space. Lastly, the demand for each node is provided. We chose to use only 7 of the 14 problems because the other problems use the exact locations and demand but with additional constraints for different variants of the VRP. The first five problems have a unique and increasingly large number of nodes. Problems 11 and 12 were included because the nodes are clustered, whereas the other problems have more evenly distributed nodes. A visual representation of problems 1 and 11 can be seen in Fig. \ref{fig:cmt1visualization} and \ref{fig:cmt11visualization}.

The experimentation was done by running the problems through DBSS, then SPS, and finally, the TS and TS with SO algorithms. For DBSS and SPS, the required number of vehicles at the start of the algorithm had to be set much higher than the actual number of vehicles to solve the problem. The number was chosen firstly to avoid default recursion depth limits in Python and, secondly, to optimize for cost. For TS and TS with SO the number of vehicles was set one number higher than was used in the BKS. This number was used so that our simple initial solution generation would always work. Table \ref{tab:vehicles} shows the vehicles used in each experiment. In reporting our experimental results, we show the total distance of the best solution found by each algorithm. We also include the deviation of that total distance with the total distance found by the BKS. Each algorithm was run three times and the best result from the three attempts is provided here. More runs would be preferred, but QA resources were limited for this research. We also include results from \cite{feld2019hybrid}. Results are shown in Table \ref{results1}. For the CMT5 problem with TS and TS w/ SO, the one-hour time limit was used as the stopping condition.

\begin{table*}[t]
    \caption{The number of vehicles used for each problem}
    \label{tab:vehicles}
    \centering
        \scalebox{0.9}{
        \begin{tabular}{cccccccccc}
            \multicolumn{2}{c}{} & \multicolumn{1}{c}{\textbf{ }} & \multicolumn{2}{c}{\textbf{DBSS}} & \multicolumn{2}{c}{\textbf{SPS}} & \multicolumn{2}{c}{\textbf{TS}} \\        
        \hline
             & \textbf{Problem} & \textbf{Vehicles used in BKS} & \textbf{Initial} & \textbf{Solution} & \textbf{Initial} & \textbf{Solution} & \textbf{Initial} & \textbf{Solution} \\
        \hline
             & CMT 1 & 5 & 6 & 6 & 6 & 6 & 6 & 5 \\
             & CMT 2 & 10 & 29 & 29 & 29 & 12 & 11 & 10 \\
             & CMT 3 & 8 & 9 & 9 & 9 & 9 & 9 & 8 \\
             & CMT 4 & 12 & 31 & 31 & 31 & 13 & 13 & 12 \\
             & CMT 5 & 17 & 43 & 43 & 43 & 19 & 18 & 17 \\
             & CMT 11 & 7 & 8 & 8 & 8 & 8 & 8 & 7 \\
             & CMT 12 & 10 & 10 & 10 & 10 & 10 & 11 & 10 \\
        \end{tabular}
        }
\end{table*}
\begin{table*}[t]
    \caption{Performance Comparison of DBScan, SPS, and TS on the CMT Dataset}
    \label{results1}
    \centering
        \scalebox{0.65}{
        \begin{tabular}{cccccccccccccc}
        \multicolumn{2}{c}{} & \multicolumn{1}{c}{\textbf{ }} & \multicolumn{1}{c}{\textbf{  }} & \multicolumn{2}{c}{\textbf{Feld at el Ref. \cite{feld2019hybrid}}} & \multicolumn{2}{c}{\textbf{DBSS}} & \multicolumn{2}{c}{\textbf{SPS}} & \multicolumn{2}{c}{\textbf{TS}} & \multicolumn{2}{c}{\textbf{TS w/ SO}} \\
        \hline
        & \textbf{Problem} & \textbf{Size} & \textbf{BKS} & \textbf{Distance} & \textbf{Dev.} & \textbf{Distance} & \textbf{Dev.} & \textbf{Distance} & \textbf{Dev.} & \textbf{Distance} & \textbf{Dev.} & \textbf{Distance} & \textbf{Dev.}\\
        \hline
         & CMT 1 & 50 & 524.61 & 556 & 5.98\% & 705 & 34.39\% & 699 & 33.24\% & 537 & 2.38\% & \textbf{524.61} & \textbf{0.0\%}\\
         & CMT 2 & 75 & 835.26 & 926 & 10.86\% & 1856 & 122.2\% & 1001 & 19.84\% & 890 & 6.62\% & \textbf{856} & \textbf{2.52\%}\\
         & CMT 3 & 100 & 826.14 & 905 & 9.55\% & 1080 & 30.73\% & 988 & 19.59\% & 938 & 13.6\% & \textbf{876} & \textbf{6.06\%}\\
         & CMT 4 & 150 & 1028.42 & 1148 & 11.63\% & 2185 & 112.46\% & 1208 & 17.46\% & 1254 & 22.01\% & \textbf{1094} & \textbf{6.4\%}\\
         & CMT 5 & 199 & 1291.29 & \textbf{1429} & \textbf{10.66\%} & 2789 & 115.99\% & 1613 & 24.91\% & 1554* & 20.41\% & 1442* & 11.72\%\\
         & CMT 11 & 120 & 1042.12 & \textbf{1084} & \textbf{4.02\%} & 1172 & 12.46\% & 1134 & 8.82\% & 1425 & 36.83\% & 1096 & 5.19\%\\
         & CMT 12 & 100 & 819.56 & 828 & 1.03\% & \textbf{827} & \textbf{0.91\%} & 876 & 6.89\% & 850 & 3.78\%  & 829 & 1.16\%\\
    \end{tabular}
    }
\end{table*}

\begin{table*}[t]
    \caption{Performance Comparison of well-known heuristics and TS on the CMT Dataset}
    \label{results2}
    \centering
        \scalebox{0.75}{
        \begin{tabular}{cccccccccccc}
        \multicolumn{2}{c}{} & \multicolumn{1}{c}{} & \multicolumn{1}{c}{} & \multicolumn{2}{c}{\textbf{Clarke-Wright}} & \multicolumn{2}{c}{\textbf{Fisher-Jaikumar}} & \multicolumn{2}{c}{\textbf{Sweep}} & \multicolumn{2}{c}{\textbf{TS w/ SO}} \\
        \hline
        & \textbf{Problem} & \textbf{Size} & \textbf{BKS} & \textbf{Distance} & \textbf{Dev.} & \textbf{Distance} & \textbf{Dev.} & \textbf{Distance} & \textbf{Dev.} & \textbf{Distance} & \textbf{Dev.} \\
        \hline
         & CMT 1 & 50 & 524.61 & 585 & 11.5\% & 524 & 0.12\% & 532 & 1.41\% & \textbf{524.61} & \textbf{0.0\%}\\
         & CMT 2 & 75 & 835.26 & 900 & 7.75\% & 857 & 2.6\% & 874 & 4.64\% & \textbf{856} & \textbf{2.52\%}\\
         & CMT 3 & 100 & 826.14 & 886 & 7.25\% & \textbf{833} & \textbf{0.83\%} & 851 & 3.01\% & 876 & 6.06\%\\
         & CMT 4 & 150 & 1028.42 & 1204 & 17.07\% & - & - & \textbf{1079} & \textbf{4.92\%} & 1094 & 6.4\%\\
         & CMT 5 & 199 & 1291.29 & 1540 & 19.26\% & 1420 & 9.97\% & \textbf{1389} & \textbf{7.57\%} & 1442* & 11.72\%\\
         & CMT 12 & 100 & 819.56 & 877 & 7.01\% & 848 & 3.47\% & 949 & 15.79\% & \textbf{829} & \textbf{1.16\%}\\
    \end{tabular}
    }
\end{table*}

\begin{table*}[t]
    \caption{Performance Comparison of DBScan, SPS, and TS on the DWave Open-Source Dataset}
    \label{results3}
    \centering
        \begin{tabular}{ccccccc}
        \multicolumn{4}{c}{} & \multicolumn{1}{c}{\textbf{ DBSS}} & \multicolumn{1}{c}{\textbf{SPS}} & \multicolumn{1}{c}{\textbf{TS w/ SO}} \\
        \hline
        & \textbf{Problem} & \textbf{Size} & \textbf{Vehicles} & \textbf{Distance}  & \textbf{Distance} & \textbf{Distance} \\
        \hline
         & Small 1 & 10 & 3 & 503 & 721 & \textbf{412} \\
         & Small 2 & 25 & 5 & 747 & 730 & \textbf{534}  \\
         & Small 3 & 50 & 7 & 1375 & 1115 & \textbf{983}  \\
         & Medium 1 & 100 & 13 & 2858 & \textbf{2124} & 2297 \\
         & Medium 2 & 150 & 15 & 3636 & \textbf{2567} & 2985 \\
         & Medium 3 & 200 & 23 & 4731 & \textbf{2965} & 3900* \\
    \end{tabular}
\end{table*}

Analysis of these results provides several areas of note. Firstly, the cost values produced by DBSS and SPS are above the BKS. Secondly, SPS provides a lower-cost solution than DBSS in almost every case. However, the cost differential between the two algorithms grows as the problem size increases. This observation does not hold, however, for the clustered datasets. In one case, the SPS algorithm performed worse; in the other, it had only a slight advantage. As we look at the results of our method we can see SO improved our results universally. Also, it is essential to note that both TS methods performed well, and TS with SO achieved the BKS for CMT1.  

We also show results in Table \ref{results2} on the CMT dataset comparing TS with SO and the following heuristics: Clarke-Wright \cite{clarke1964scheduling}, Fisher-Jaikumar \cite{fisher1981generalized}, and Sweep \cite{gillett1974heuristic}. The heuristic results were compiled from \cite{feld2019hybrid} and are here to compare TS capabilities.

The analysis is consistent with our results compared to other hybrid algorithms in that our method can sometimes outperform well-known heuristics. Interestingly, TS consistently outperforms Clarke-Wright's savings algorithm \cite{clarke1964scheduling}.  

\begin{figure}[hbtp]
    \centering
    \fbox{\includegraphics[width=\linewidth]{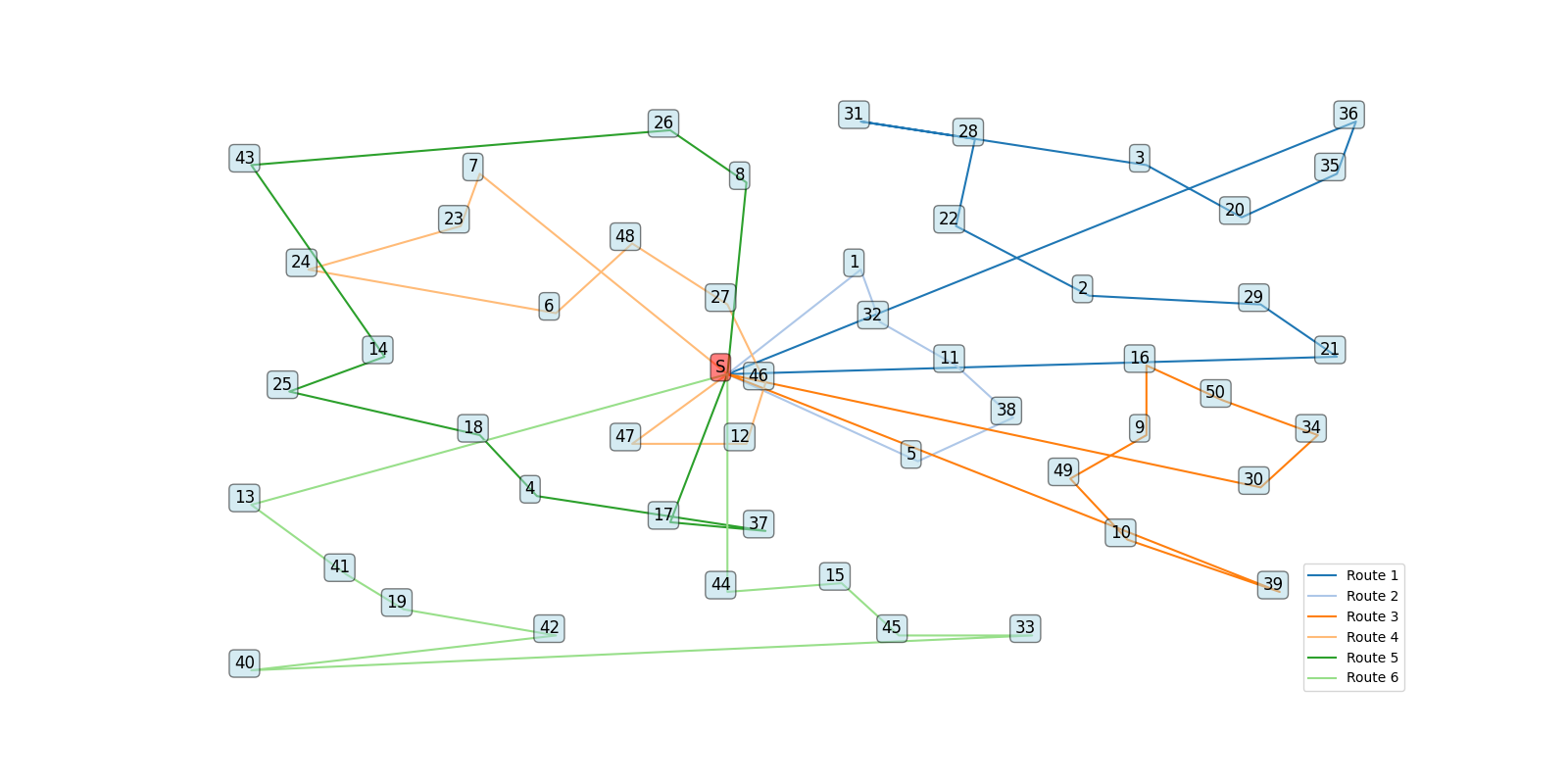}}
    \caption{CMT 1 Result using DBSS}
    \label{fig:mesh1}
\end{figure}

\begin{figure}[hbtp]
    \fbox{\includegraphics[width=\linewidth]{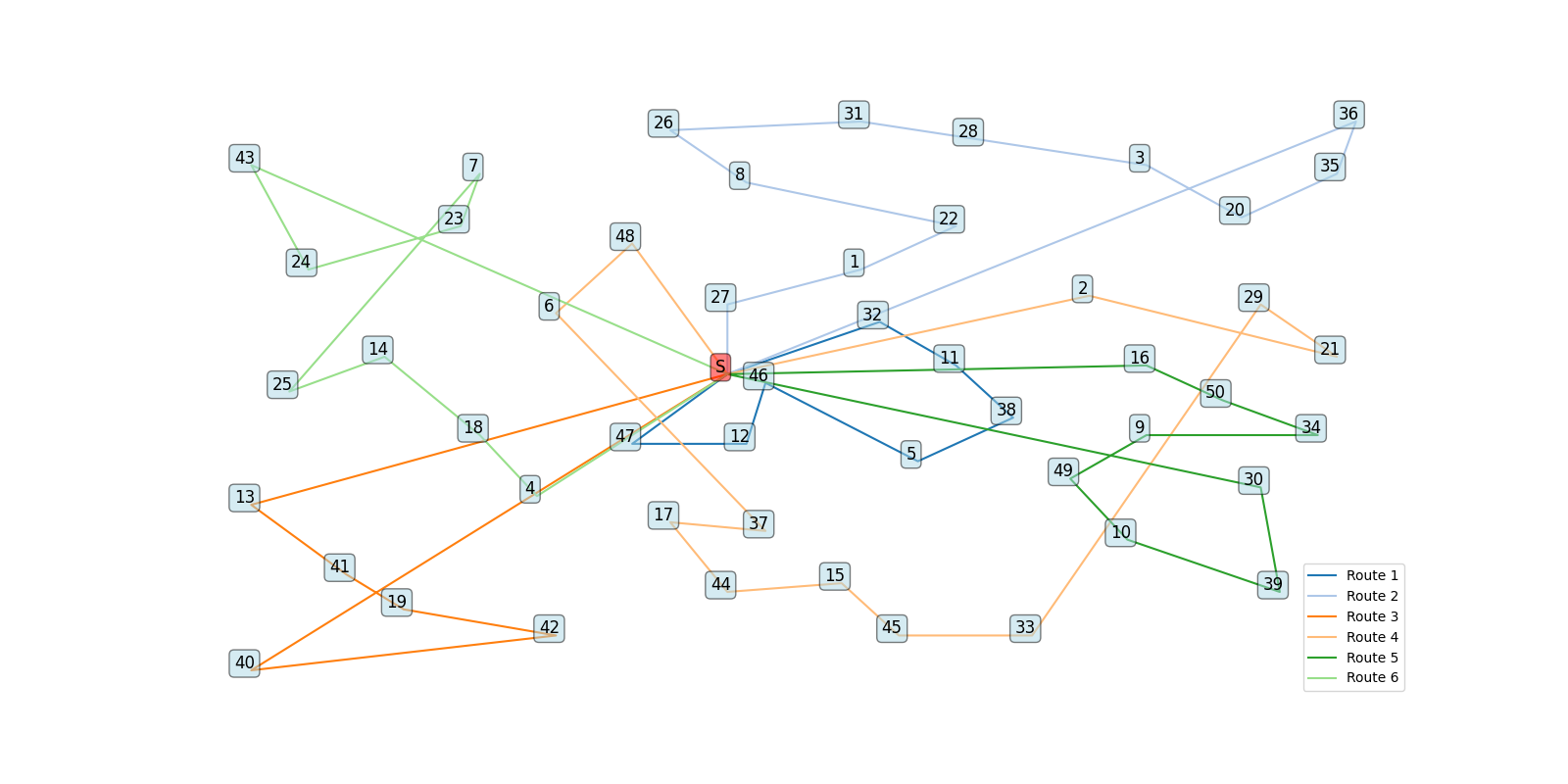}}
    \caption{CMT 1 Result using SPS}
    \label{fig:mesh2}
\end{figure}

\begin{figure}[hbtp]
    \fbox{\includegraphics[width=\linewidth]{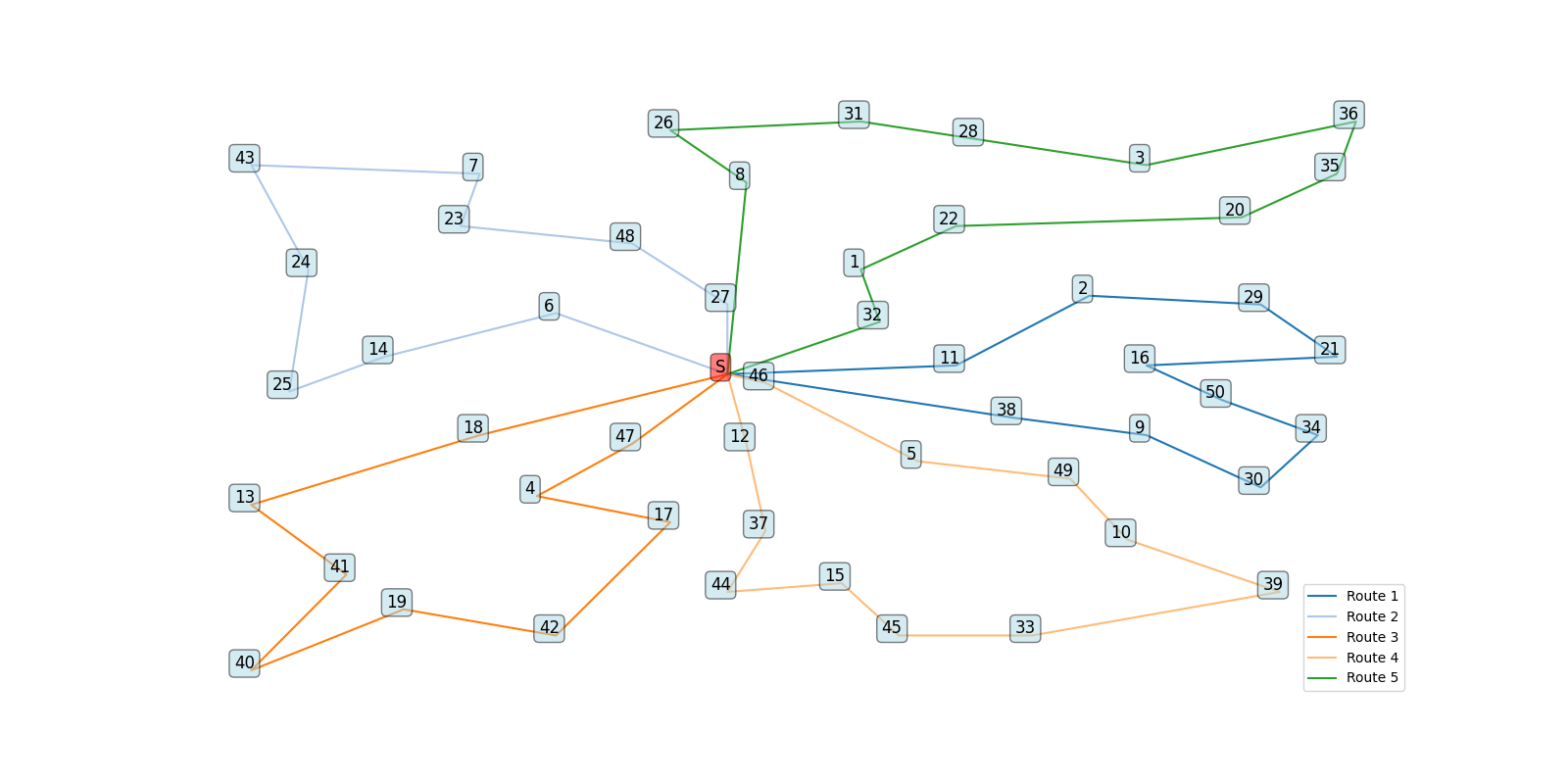}}
    \caption{CMT 1 Result using TS w/ SO}
    \label{fig:mesh3}
\end{figure}

Figures \ref{fig:mesh1}, \ref{fig:mesh2}, and \ref{fig:mesh3} visualize the routes created by DBSS, SPS, and TS with SO for the CMT 1 problem. TS here is showing the BKS. It is noted in the BKS that none of the routes cross any other routes, and there are no crossed lines within a route either. These are hallmark features of good routes and a good overall solution.  


\subsection{DWAVE Dataset}

The second dataset we evaluated was referenced by \cite{borowski2020new} and provided by \cite{githubdwave}. This dataset includes 51 problems split into four categories: small, medium, big, and mixed. We tested the small and medium problems 1-3. This dataset presents a couple of modifications to the VRP. Here, the fleet is heterogeneous because each vehicle has a different capacity. Also, these problems allow for multiple source and sink nodes in contrast with the CMT dataset. Thus, the problem in this dataset is categorized as the multi-depot heterogeneous fleet CVRP (MDHFVRP). The dataset does not provide coordinates for the nodes but instead gives the cost between two nodes. The vehicle number was not modified from the values that were given. The results are shown in Table \ref{results3}. For the Medium 3 problem with TS w/ SO, the one-hour time limit was used as the stopping condition.

Analysing the results, we see that SPS does not consistently outperform DBSS. In this case, it is clear that as the tests get larger, the differential in the advantage of SPS over DBSS grows. However, if the test is sufficiently small, then DBSS outperforms SPS. For this dataset, we only evaluated the TS with SO. For the small problems, TS could consistently outperform the algorithms SPS and DBSS but could not outperform the medium problems. We concluded that the neighborhood mechanism we utilized to limit our local search was counterproductive with this dataset. Routes that start and end at different locations no longer always share a property of needing the locations on the route to be clustered close together. It also explains why DBSS is performing poorly on this dataset as well. The time cost to consider every possible move was too large as we moved to the medium problems. A more strategic local search would provide faster convergence and an even better resulting algorithm.

\section{Conclusions} \label{Sec:Concl}
\begin{figure}[t]
    \centering
    \includegraphics[width=\linewidth]{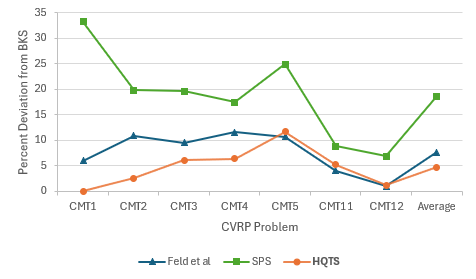}
    \caption{Percent deviation from BKS for three hybrid algorithms}
    \label{fig:deviation}
\end{figure}

In this paper, we have demonstrated that HQTS (TS and TS with SO) can solve the CVRP and stand with other hybrid algorithms. Our results, in many cases, outperform other hybrid algorithms, and we even found the BKS in one case. In fact, our method, on average, across all of the CMT problems, deviated the least from the BKS Fig. \ref{fig:deviation}. While we successfully showcased quantum computing inside a metaheuristic, our simple approach to TS limited our overall effectiveness. While that was somewhat expected because we set out to implement a simple TS approach to showcase the power of quantum computing. Still, our limited access to QA forced us to implement a slower-performing algorithm. For our future work, we plan to improve our access to QA or utilize simulated QA to create a more efficient TS. Also, we plan to refine our approach further so that HQTS will consistently find the BKS. The challenge we discovered with the MDHFVRP variant of the VRP has inspired us to look for a more robust TS to solve this problem. It aligns with our future goal of solving more highly constrained variants of the VRP that are common in the industry.

{
    \small
    \bibliographystyle{IEEEtran}
    \bibliography{references}

\begin{thebibliography}{10}
\providecommand{\url}[1]{#1}
\csname url@samestyle\endcsname
\providecommand{\newblock}{\relax}
\providecommand{\bibinfo}[2]{#2}
\providecommand{\BIBentrySTDinterwordspacing}{\spaceskip=0pt\relax}
\providecommand{\BIBentryALTinterwordstretchfactor}{4}
\providecommand{\BIBentryALTinterwordspacing}{\spaceskip=\fontdimen2\font plus
\BIBentryALTinterwordstretchfactor\fontdimen3\font minus \fontdimen4\font\relax}
\providecommand{\BIBforeignlanguage}[2]{{%
\expandafter\ifx\csname l@#1\endcsname\relax
\typeout{** WARNING: IEEEtran.bst: No hyphenation pattern has been}%
\typeout{** loaded for the language `#1'. Using the pattern for}%
\typeout{** the default language instead.}%
\else
\language=\csname l@#1\endcsname
\fi
#2}}
\providecommand{\BIBdecl}{\relax}
\BIBdecl

\bibitem{dantzig1954solution}
G.~Dantzig, R.~Fulkerson, and S.~Johnson, ``Solution of a large-scale traveling-salesman problem,'' \emph{Journal of the operations research society of America}, vol.~2, no.~4, pp. 393--410, 1954.

\bibitem{eksioglu2009vehicle}
B.~Eksioglu, A.~V. Vural, and A.~Reisman, ``The vehicle routing problem: A taxonomic review,'' \emph{Computers \& Industrial Engineering}, vol.~57, no.~4, pp. 1472--1483, 2009.

\bibitem{clarke1964scheduling}
G.~Clarke and J.~W. Wright, ``Scheduling of vehicles from a central depot to a number of delivery points,'' \emph{Operations research}, vol.~12, no.~4, pp. 568--581, 1964.

\bibitem{ladd2010quantum}
T.~D. Ladd, F.~Jelezko, R.~Laflamme, Y.~Nakamura, C.~Monroe, and J.~L. O’Brien, ``Quantum computers,'' \emph{nature}, vol. 464, no. 7285, pp. 45--53, 2010.

\bibitem{laporte2013vehicle}
G.~Laporte, P.~Toth, and D.~Vigo, ``Vehicle routing: historical perspective and recent contributions,'' \emph{EURO Journal on Transportation and Logistics}, vol.~2, pp. 1--4, 2013.

\bibitem{irie2019quantum}
H.~Irie, G.~Wongpaisarnsin, M.~Terabe, A.~Miki, and S.~Taguchi, ``Quantum annealing of vehicle routing problem with time, state and capacity,'' in \emph{Quantum Technology and Optimization Problems: First International Workshop, QTOP 2019, Munich, Germany, March 18, 2019, Proceedings 1}.\hskip 1em plus 0.5em minus 0.4em\relax Springer, 2019, pp. 145--156.

\bibitem{feld2019hybrid}
S.~Feld, C.~Roch, T.~Gabor, C.~Seidel, F.~Neukart, I.~Galter, W.~Mauerer, and C.~Linnhoff-Popien, ``A hybrid solution method for the capacitated vehicle routing problem using a quantum annealer,'' \emph{Frontiers in ICT}, vol.~6, p.~13, 2019.

\bibitem{osaba2022systematic}
E.~Osaba, E.~Villar-Rodriguez, and I.~Oregi, ``A systematic literature review of quantum computing for routing problems,'' \emph{IEEE Access}, vol.~10, pp. 55\,805--55\,817, 2022.

\bibitem{dwave2024}
\BIBentryALTinterwordspacing
D.-W.~S. Inc, ``What is quantum annealing?'' 2024, accessed on February 02 2024. [Online]. Available: \url{https://docs.dwavesys.com/481 docs/latest/c\_gs\_2.html}
\BIBentrySTDinterwordspacing

\bibitem{farhi2000quantum}
E.~Farhi, J.~Goldstone, S.~Gutmann, and M.~Sipser, ``Quantum computation by adiabatic evolution,'' \emph{arXiv preprint quant-ph/0001106}, 2000.

\bibitem{harwood2021formulating}
S.~Harwood, C.~Gambella, D.~Trenev, A.~Simonetto, D.~Bernal, and D.~Greenberg, ``Formulating and solving routing problems on quantum computers,'' \emph{IEEE Transactions on Quantum Engineering}, vol.~2, pp. 1--17, 2021.

\bibitem{fitzek2021applying}
D.~Fitzek, T.~Ghandriz, L.~Laine, M.~Granath, and A.~F. Kockum, ``Applying quantum approximate optimization to the heterogeneous vehicle routing problem,'' \emph{arXiv preprint arXiv:2110.06799}, 2021.

\bibitem{matsubara2020digital}
S.~Matsubara, M.~Takatsu, T.~Miyazawa, T.~Shibasaki, Y.~Watanabe, K.~Takemoto, and H.~Tamura, ``Digital annealer for high-speed solving of combinatorial optimization problems and its applications,'' in \emph{2020 25th Asia and South Pacific Design Automation Conference (ASP-DAC)}.\hskip 1em plus 0.5em minus 0.4em\relax IEEE, 2020, pp. 667--672.

\bibitem{harikrishnakumar2020quantum}
R.~Harikrishnakumar, S.~Nannapaneni, N.~H. Nguyen, J.~E. Steck, and E.~C. Behrman, ``A quantum annealing approach for dynamic multi-depot capacitated vehicle routing problem,'' \emph{arXiv preprint arXiv:2005.12478}, 2020.

\bibitem{yarkoni2021solving}
S.~Yarkoni, A.~Huck, H.~Sch{\"u}lldorf, B.~Speitkamp, M.~S. Tabrizi, M.~Leib, T.~B{\"a}ck, and F.~Neukart, ``Solving the shipment rerouting problem with quantum optimization techniques,'' in \emph{Computational Logistics: 12th International Conference, ICCL 2021, Enschede, The Netherlands, September 27--29, 2021, Proceedings 12}.\hskip 1em plus 0.5em minus 0.4em\relax Springer, 2021, pp. 502--517.

\bibitem{bao2021approach}
S.~Bao, M.~Tawada, S.~Tanaka, and N.~Togawa, ``An approach to the vehicle routing problem with balanced pick-up using ising machines,'' in \emph{2021 International Symposium on VLSI Design, Automation and Test (VLSI-DAT)}.\hskip 1em plus 0.5em minus 0.4em\relax IEEE, 2021, pp. 1--4.

\bibitem{bao2023ising}
------, ``An ising-machine-based solver of vehicle routing problem with balanced pick-up,'' \emph{IEEE Transactions on Consumer Electronics}, 2023.

\bibitem{ajagekar2020quantum}
A.~Ajagekar, T.~Humble, and F.~You, ``Quantum computing based hybrid solution strategies for large-scale discrete-continuous optimization problems,'' \emph{Computers \& Chemical Engineering}, vol. 132, p. 106630, 2020.

\bibitem{lucas2014ising}
A.~Lucas, ``Ising formulations of many np problems,'' \emph{Frontiers in physics}, vol.~2, p. 74887, 2014.

\bibitem{borowski2020new}
M.~Borowski, P.~Gora, K.~Karnas, M.~B{\l}ajda, K.~Kr{\'o}l, A.~Matyjasek, D.~Burczyk, M.~Szewczyk, and M.~Kutwin, ``New hybrid quantum annealing algorithms for solving vehicle routing problem,'' in \emph{International Conference on Computational Science}.\hskip 1em plus 0.5em minus 0.4em\relax Springer, 2020, pp. 546--561.

\bibitem{christofides1981exact}
N.~Christofides, A.~Mingozzi, and P.~Toth, ``Exact algorithms for the vehicle routing problem, based on spanning tree and shortest path relaxations,'' \emph{Mathematical programming}, vol.~20, pp. 255--282, 1981.

\bibitem{gillett1974heuristic}
B.~E. Gillett and L.~R. Miller, ``A heuristic algorithm for the vehicle-dispatch problem,'' \emph{Operations research}, vol.~22, no.~2, pp. 340--349, 1974.

\bibitem{fisher1981generalized}
M.~L. Fisher and R.~Jaikumar, ``A generalized assignment heuristic for vehicle routing,'' \emph{Networks}, vol.~11, no.~2, pp. 109--124, 1981.

\bibitem{cordeau2002guide}
J.-F. Cordeau, M.~Gendreau, G.~Laporte, J.-Y. Potvin, and F.~Semet, ``A guide to vehicle routing heuristics,'' \emph{Journal of the Operational Research society}, vol.~53, pp. 512--522, 2002.

\bibitem{osaba2021hybrid}
E.~Osaba, E.~Villar-Rodriguez, I.~Oregi, and A.~Moreno-Fernandez-de Leceta, ``Hybrid quantum computing-tabu search algorithm for partitioning problems: preliminary study on the traveling salesman problem,'' in \emph{2021 IEEE Congress on Evolutionary Computation (CEC)}.\hskip 1em plus 0.5em minus 0.4em\relax IEEE, 2021, pp. 351--358.

\bibitem{glover1989tabu}
F.~Glover, ``Tabu search—part i,'' \emph{ORSA Journal on computing}, vol.~1, no.~3, pp. 190--206, 1989.

\bibitem{glover1990tabu}
------, ``Tabu search—part ii,'' \emph{ORSA Journal on computing}, vol.~2, no.~1, pp. 4--32, 1990.

\bibitem{osman1993metastrategy}
I.~H. Osman, ``Metastrategy simulated annealing and tabu search algorithms for the vehicle routing problem,'' \emph{Annals of operations research}, vol.~41, pp. 421--451, 1993.

\bibitem{taillard1993parallel}
{\'E}.~Taillard, ``Parallel iterative search methods for vehicle routing problems,'' \emph{Networks}, vol.~23, no.~8, pp. 661--673, 1993.

\bibitem{gendreau1994tabu}
M.~Gendreau, A.~Hertz, and G.~Laporte, ``A tabu search heuristic for the vehicle routing problem,'' \emph{Management science}, vol.~40, no.~10, pp. 1276--1290, 1994.

\bibitem{gendreau1992new}
------, ``New insertion and postoptimization procedures for the traveling salesman problem,'' \emph{Operations Research}, vol.~40, no.~6, pp. 1086--1094, 1992.

\bibitem{glover2011case}
F.~Glover and J.-K. Hao, ``The case for strategic oscillation,'' \emph{Annals of Operations Research}, vol. 183, pp. 163--173, 2011.

\bibitem{githubdwave}
\BIBentryALTinterwordspacing
D.-W.~S. Inc, ``An approach to solve the vehicle routing problem (vrp) using quantum computing,'' 2023, accessed on January 15 2024. [Online]. Available: \url{https://github.com/dwave-examples/D-Wave-VRP}
\BIBentrySTDinterwordspacing

\end{thebibliography}
}

\end{document}